\documentclass{article}

\newcommand{\address}[1]{\gdef\@address{#1}}%
\usepackage{cite}
\usepackage{sansmath}
\usepackage{authblk}

\usepackage{url}
\def\thetitle{Multi-scale and multi-domain computational astrophysics}

\begin{document}

\pagestyle{myheadings}
\title{\thetitle}

\author[*]{Arjen van Elteren}
\author[*]{Inti Pelupessy}
\author[* $\dag$]{Simon Portegies Zwart}

\affil[*]{
Leiden Observatory, Leiden University, PO Box 9513, 2300 RA, Leiden, The Netherlands
}

\affil[$\dag$]{E-mail: spz@strw.leidenuniv.nl}
%%%% Subject entries to be placed here %%%%
%\subject{xxxxx, xxxxx, xxxx}

%%%% Keyword entries to be placed here %%%%
%\keywords{xxxx, xxxx, xxxx}

\maketitle

\begin{abstract}
Astronomical phenomena are governed by processes on all spatial and
temporal scales, ranging from days to the age of the Universe
(13.8\,Gyr) as well as from km size up to the size of the Universe.
This enormous range in scales is contrived, but as long as there is a
physical connection between the smallest and largest scales it is
important to be able to resolve them all, and for the study of many
astronomical phenomena this governance is present. Although covering
all these scales is a challenge for numerical modelers, the most challenging
aspect is the equally broad and complex range in physics, and the way
in which these processes propagate through all scales.  In our recent
effort to cover all scales and all relevant physical processes on
these scales we have designed the Astrophysics Multipurpose Software
Environment (AMUSE). AMUSE is a Python-based framework with production
quality community codes and provides a specialized environment to
connect this plethora of solvers to a homogeneous problem solving
environment.
\end{abstract}

\footnotetext[0]{Submitted to: Philosophical Transactions A}
%\begin{fmtext}
%\keywords{}
%\end{fmtext}

%%%%%%%%%%%%%%% End of first page %%%%%%%%%%%%%%%%%%%%%

%\maketitle

\section{Introduction}

The universe evolves over multiple scales in time and space.  This wide 
range in scales is generally hard to simulate on a standard 64-bit 
computer. If a causal connection between the scales exists  
multi-scale simulations are necessary.  In some cases, such a 
connection is made, for example in gravity, where the long-range and 
the absence of shielding causes the small scales to be connected with 
the largest scales. To the extreme, one can argue that the dynamical 
evolution of a star at the other side of the Milky Way Galaxy may 
affect life on the planet Earth. This works as follows: the star at the 
other side of the Milky Way may have been in the cluster in which the 
Solar system was born. Early perturbations in the solar system may than 
have originated from this nearby star, which in the last 4.6\,Gyr has 
moved away to be at the other side of the Milky Way today. On the same 
account, a star that is in the neighborhood today may have been born in 
an entirely different galaxy, which merged several Gyr ago with the 
Milky Way Galaxy. In this way long ago, and far away events may have a 
direct imprint on the evolution of the solar system here and now, and 
with that on the evolution of the planet Earth.  In order to appreciate 
the complexity of the dynamical evolution of the Galaxy scales ranging 
in size from one to $10^{18}$\,km ($10^5$\,parsec), and from a single 
second to over $10^{17}$\,seconds (4.6\,Gyr) should be appreciated.

Incorporating such a wide range in scales in numerical simulations is
extremely challenging, and luckily for many modern studies most scales
can be omitted with the argument that the relative importance of
mutual interactions between the smallest and the largest scales is
small. In other cases this decoupling is not so evident, and such an 
argument should be verified explicitly.

On top of the different time and distance scales in our Universe, the 
physical processes happening on these scales have traditionally been 
split into different domains. Gravity underpins most of the processes 
in the Universe, and models have been developed to predict the typical 
evolution of gravitational bodies at different scales. From very 
specialized models that can predict the evolution of a small set of 
bodies \((N<10^2)\) \cite{1999MNRAS.304..793C} 
\cite{2008AJ....135.2398M}, and models that can accurately predict the 
dynamics of a modest number of bodies \((10^2<N<10^4)\) 
\cite{2007NewA...12..357H} \cite{1999PASP..111.1333A} to models that 
are less accurate but able to predict the overall evolution of many 
bodies \((10^6<N<10^9)\) \cite{2005MNRAS.364.1105S} 
\cite{2012JCoPh.231.2825B}. Star clusters, stars and planets are 
created from collapsing clouds of gas on different length and time 
scales. Like gravitational models, many hydrodynamical models for 
predicting the evolution of these gas clouds have been developed 
\cite{2008ApJS..178..137S} \cite{1989ApJS...70..419H} 
\cite{1991A&A...252..718M}, each having unique properties that we would 
like to utilize. Similarly, to describe the evolution of a single star, 
a variation of stellar evolution models can be applied. From plasma 
magneto-hydrodynamical models for very short time and distance scales 
\cite{2012JCoPh.231..718K}, and 1d spherical symmetrical models of a 
star for an accurate description of the processes inside a star over 
long time-scale \cite{1971MNRAS.151..351E} \cite{2008A&A...488.1007G} 
\cite{2011ApJS..192....3P}, to analytically models that describe a few 
important variable of a star over very long time scales 
\cite{2000MNRAS.315..543H} \cite{1996A&A...309..179P}. Finally, 
radiative processes need to be modeled to predict the ionization 
levels and energy balance of the different gas clouds in the Universe. 
Specialized models exists for radiation on different bands like X-ray 
and UV, different time-scales and space scales 
\cite{2010A&A...515A..79P} \cite{2008MNRAS.386.1931A}. 

In the past 4 years we have developed the Astrophysics Multipurpose 
Software Environment (AMUSE) to simulate parts of our Universe. Within 
AMUSE we want to include large collection of codes, to be agnostic to 
which implementation of a physical model we can to use for any 
simulation. Also, apart from the physical model, the actual algorithm 
often plays an important role in the accuracy and efficiency by which 
we can simulate that physical model. Finally, a large part of our 
predictive research is figuring out which physical processes play an 
important role in the system we are interested in and we do not want to 
be pinned to any specific model or algorithm. 

We have designed AMUSE as a toolbox of different simulation codes with
specialized support to couple these codes based on different scales
and physics \cite{2013CoPhC.183..456P} \cite{2009NewA...14..369P}
\cite{2012ASPC..453..317P} \cite{2008AN....329..885H}. In this paper
we will discuss our experiences in the development of AMUSE, which is
a framework for multi-scale and multi-domain physics. AMUSE is
publicly available at {\tt http://amusecode.org}.

\section{Unit conversion}

Most scientific codes operate in an implicit and predetermined set of 
units which are usually code-specific. The units are rarely documented 
and the user is confronted with the implicitly adopted usage of unit 
values by the implementer.  For example, in the evolution of self 
gravitating systems, the dimensionless N-body units 
\cite{1986LNP...267..233H} are very common. Though easy to understand 
for the expert, because they allow the results to scale with respect to 
the dimensionless parameters, they are hard to understand for the 
novice. So long as a code is mono-physics the implicit assumption of 
units is a fine choice. However, units are essential for long term 
maintenance, to enable inter-code operation and for the understanding 
of novice users.

All values of variables, attributes or function parameters are 
explicitly quantified with a unit. These values are implemented as 
\emph{Quantity} objects and all interactions with the codes is via 
these objects. In many ways, using quantities is like specifying value 
types. In this case, numbers are part of a set with a fixed collection 
of arithmetic operators for that set. The same is true for the 
quantities, only now the arithmetic operators check and convert the 
units with the right semantics for the given operator. For physical 
simulations, where almost all values are rational, we find the use of 
quantities more powerful that value types. The units of the values 
in our models differ more than the value types.

The use of \emph{Quantities} and \emph{Units} throughout the code has 
turned out to be beneficial in a number of ways outside of their 
intended use to enable coupling between codes. A user does not have to 
familiarize herself with the specific units or scaled units used in 
each physics domain, removing a real barrier in learning to use a code. 
As units can be constructed from other units with arithmetic operators, 
the user can extend these to fit the problem under investigation. 
Finally, all units and quantities are checked for each operation and the 
use of units has assisted us in preventing bugs. When a user assigns a 
mass to a radius, an error is raised exactly specifying the problem, 
which would be much harder to detect if propagated through the 
simulation.
  
\section{Data representation}

All scientific codes manipulate and store evolving data values on the 
system being modeled. How this data is represented in a code is often 
demanded by the algorithms applied in that code or determined 
by previous experiences of the implementer. We found a great variety of 
data representations: flat tables, various tree structures, multi 
dimensional arrays, complex structures or combinations of these. 
Although optimal for the specific implementation, exposing such a large 
collection of data representations would impede inter-code operation 
and understanding of the framework.

We have implemented two different data representations. \emph{Particle 
sets} provide object based access to the code data. Objects can be 
added or removed and attribute values can be requested and updated on 
the objects. \emph{Grids} provide multi-dimensional indexed access to 
the code data, a grid index space is fixed after creation, attribute 
values can be requested and updated on the grid points. The definition 
of the attributes can be changed per code and are often specific per 
physics domain. Linking between the objects or grid points in the data 
representation is possible, but these links are supposed to be 
physically meaningful and not numerical. The relation between a binary 
particle and the two constituent stars belonging to it is exposed, 
whereas the tree structure of an AMR code is not. 

Common operations are simplified by having only two data 
representations. One initial condition creation method can be used for 
multiple codes. If codes have compatible data representations, data 
exchange between codes can be generalized. Reporting functionality can 
be reused across codes. 

\section{Evolution of a system}

Usually, one is interested in the evolution of a system forward in time. 
How the evolution of a system is divided over the codes is not 
constrained within AMUSE, but the decision of the researcher. Although 
not strictly necessary, usually the entire state of a system is 
synchronized at prescribed time intervals. For the timestepping of the 
code interactions a constant global timestep can be used (this is the 
simplest case), but more complicated schemes are possible (although 
progressively more difficult to implement). For the evolution of 
subsystems, the codes themselves are free to choose their internal 
timestepping scheme (this is invisible from the framework). We have 
identified two recurrent patterns to divide the evolution over codes, 
parallel and sequential:

In most coupled problems, codes will update the state of the system in 
parallel. Each code evolves their piece of the system to the next point 
in time of the whole system. In this kind of coupling, codes update 
different attribute values of the same object or distinct objects are 
updated. In modeling a stellar cluster, a gravitational code will 
evolve the positions and velocities of the stars, while a stellar 
evolution code will evolve the mass and radius of the same stars. A 
hydrodynamics code, for example, will evolve an embedded gas cloud 
separately but in lockstep with the gravity and the stellar evolution 
code.

Other problems demand a sequential evolution of the system by different 
codes. A single code can generally not evolve the system to the end 
state. In a stellar collision problem, a gravitational code can 
integrate the orbits of two stars as point masses until the stars are 
close to approach each other to some pre-specified distance upon which 
an exception is raised. Just before the collision, a hydrodynamics code 
may take over the state of the system and evolve the hydrodynamic 
collision process. The resulting end state could subsequently be used 
as an initial condition of yet another code for further evaluation.

The former case of parallel couplings occurs where no temporal 
seperation exists, while the sequential pattern is applicable where a 
strong temporal separation can be made. In both cases the physical 
and/or spatial scales may be different between sub-solvers. The above 
patterns can be combined to produce more complex divisions of the 
evolution of a system. In the stellar collision problem, a stellar 
evolution code could evolve the overall structure of a star during the 
gravitational point mass phase of the simulation. This structure can 
than be added to the initial conditions for the hydrodynamics code. 

\section{Communication between codes}

Codes must be able to interact with each other to be able to evolve a 
multi-domain and multi-scale system. We have implemented this 
interaction in the controlling framework layer. This makes the method 
by which codes influence each other explicit for the researcher. The 
implementation of the coupling between codes is also exactly the point 
where our scientific research is focused. Instead of implementing a 
number of fixed coupling methods inside the codes, we have implemented 
several generic components on top of the codes. The researcher can 
combine these components to implement the required coupling based on 
the physics.

%A consistent view on the quantities, data representation and the 
%evolution in time of the codes form the basis for the coupling 
%components. 

The component that is used in almost all coupling methods is the 
\emph{Channel}. \emph{Channels} enable the researcher to exchange data 
between two codes, using the data representations of the codes. An 
important functionality of \emph{Channels} is to map the identity of 
the particles or grid points of both codes. The same star can be 
defined in a stellar evolution code and a gravitational dynamics code. 
To copy the mass of the star from the stellar evolution code to the 
gravitational dynamics code, the identity of that star in both codes 
must be ensured. \emph{Channels} keep track of the objects in both 
codes and are able to perform the correct update and handle the correct 
mapping when objects are removed or added in one or both codes. 

For other coupling methods, the evolution in a code must be interrupted 
when a special condition is encountered. This functionality is provided 
by an exception which we call a  \emph{Stopping Condition}. When a 
\emph{Stopping Condition} is flagged, the code will finish its current 
evolution step and give control back to the controlling layer. In this 
layer the particles involved in the \emph{Stopping Condition} are 
identified and processed based on the actual coupling method. A number 
of different stopping conditions exists. For hydrodynamics codes one of 
the stopping conditions flags a high density region in the system. This 
condition is used to initiate a sink and source coupling method where 
stars will form and accrete gas in these high density regions.

Some codes also implement a \emph{Service}. Physical quantities used 
for the coupling methods can be calculated by a code based on the 
current state of the model inside that code. Gravity codes can 
calculate the acceleration field at arbitrary points. Hydrodynamics 
codes can calculate the local density and momentum at arbitrary points. 
An implementation of a \emph{Service} is added to a code when importing 
this code into the framework. An interesting feature of a 
\emph{Service} is that it can be implemented in the controlling layer, 
only this is more intricate as all relevant data requires to be copied 
and some of the algorithms of the code need to be re-implemented.

\section{Multi-scale astrophysics}

To handle multi-scale problems, the physical processes on different 
scales may influence each other, but it must be possible to separately 
model this influence. The problem can than be split over different 
codes combined with one ore more of the coupling methods that implement 
the physical coupling processes. We have implemented several coupling 
methods for a variety of physics. We will demonstrate one method in 
this paper. %, both for gravitational dynamics.

In star cluster simulations most interactions are over large time and 
length scales and the mean stellar movement is determined by the 
complete gravitational field of the cluster. For the individual track 
of the stars and the evolution of the cluster as a whole, close 
encounters between a small number of particles are crucial. The scale 
on which the encounters occur is much smaller than the evolution scale 
of the entire cluster. Evolving the cluster on these small scales is 
often not practical. The \emph{Multiples} coupling method handles the 
different scales by splitting the evolution between large scale cluster 
evolution and small scale close encounters. The entire system is 
evolved sequentially between codes. Initially, the cluster is evolved 
as a set of single stars in a gravitational dynamics code.  The 
gravitational dynamics code must support a \emph{Stopping Condition} to 
flag when two stars come close together. When this \emph{Stopping 
Condition} is flagged, the two stars in the encounter are evolved in a 
second gravitational dynamics code. By the time the encounter is 
resolved, a binary may have formed or the stars are separating and 
moving on. The binary will be modeled as a single star in the first 
gravitational dynamics code and the large scale evolution of the system 
is continued. The system evolves until the next encounter, and 
after completing the handling of the next encounter the large scale 
evolution of the system is again continued. In this way the same 
gravitational processes are resolved on two very different scales.

Two conditions must be fulfilled to be able to split the processes on 
different scales. Every close encounter must resolve and this 
resolution must occur instantaneously with the respect to the large 
scale cluster evolution. Close encounters typically occur on the 
timescale of the local crossing time (in the order of years) compared 
to the global crossing time (in the order of million years) of a star 
in the cluster. Also, for the binaries to be decoupled from the large 
scale evolution of the cluster, its semi-major axis must be small 
compared to the cluster scales. Every split of a system between codes 
creates conditions like theses. Whether these conditions are fulfilled 
must always be examined for the specific problem. This examination is 
supported by all coupling methods being explicit and open for 
investigation by a researcher.

%Another interesting star cluster evolution problem, is simulating solar 
%systems surrounding the stars in an evolving cluster. Close encounters 
%may kick planets from the solar systems into the cluster. These kind of 
%systems are evolved with the \emph{Tree} coupling method. In this 
%method the small and the large system are evolved in parallel by 
%diffent codes. The small systems are modelled as single particles in 
%the large systems. Like \emph{Multiples}, when a stars come within 
%close to a small system, these will be added to the small systems. When 
%a particle escapes a small system it is added to the larger system. As 
%this process can be modelled over several scale \emph{steps} it can be 
%represented as a \emph{Tree} of increasingly smaller subsystems.

\section{Multi-domain astrophysics}

For multi-domain astrophysics, the physics of a code from one domain 
influences the physics of a code in another domain. Like the 
multi-scale approach, it must be possible to split the problem in two 
separate processes to couple the physics between the two domains.

Some of the multi-domain physics can be coupled by updating the value of 
an attribute in one code based on the evolution of that attribute in 
another code. The evolution of a star in a cluster will not be 
influenced by the gravitational field of that cluster. The mass loss 
that occurs during stellar evolution will influence the 
gravitational dynamics of the cluster, by changing the mass of a 
particle. One channel is used to transport this information from the 
stellar evolution code to the gravitational dynamics code. By changing 
the masses in one code, based on the physics in the other code we have 
implemented a trivial multi-domain physics problem.

Other multi-domain problems require careful analysis of the physics to 
be able to make a meaningful separation between the codes. The 
\emph{Bridge} coupling method \cite{2007PASJ...59.1095F} is used to 
evolve the particles of a one code inside the gravitational field of 
another code. The coupling of the gravitational force between particles 
is based on splitting of the Hamiltonian of the coupled system 
\cite{2012MNRAS.420.1503P}. With this strict separation one can define 
the Hamiltonian of both systems and two partial Hamiltonians that 
describe the influence of one system on the other and vise verse. This 
can be further simplified to a process where the particles of one code 
are influenced by the field created by the particles of other codes or 
determined by analytical models. In this way the \emph{Bridge} coupling 
method can add gravitational force to the evolution of the system in a 
code without support for this force. The hydrodynamics of an embedded 
gas cloud in a star cluster can be simulated with the field of the 
stars in the cluster and also with the field of the gas itself without 
the hydrodynamics code having an implementation of gravitational 
dynamics.

Intrinsically coupled multi-domain astrophysics, where the problem 
cannot be split into separate parts with interaction terms, are not 
supported by the coupling methods. For these problems, like 
magneto-hydrodynamics or post Newtonian gravitational dynamics, special 
codes must be added to be able to provide the coupling. Fortunately, 
codes that implement these multi-domain physics models, already exist 
and can be integrated like mono-physics codes.
 
The exact restrictions and the requirements in terms of the astrophysical 
applications of the different coupling schemes for multi-domain and 
multi-scale simulations are difficult to generalize, and remain the 
responsibility of the researcher to check. This can be done by monitoring 
diagnostics of the simulation (e.g. energy conservation), conducting 
convergence studies and validating the sub-couplings over the expected 
parameter range or by reference to test problems with known results.
 
% TODO KIRA

\section{Conclusion}

It is challenging to computationally evolve an astrophysical system
over multiple scales involving a variety of physical processes. We
have tried to tackle this problem by creating a framework of
astrophysical codes were all quantities have units, all data is
represented in the same way and all codes provide standard components
that support interaction between the codes. Since the coupling
between the codes is implemented in the controlling layer
of the framework, this approach depends on the possibility to split
the physics of a system into separate parts either over the scales or
over the type of physics involved or both. We have found that AMUSE is 
well suited for a range of problems in astrophysics where such a separation
suggests itself from te nature of the problem. In other cases, where the
physical processes are tightly intertwined it is best to import the physical
solver in question as a single code. The exact boundary between these two
cases remains an area of active research.

AMUSE is completely
free and we distribute it as open source via the website {\tt
  http://amusecode.org}.

%Problems often have multi-scale and multi-domain physical aspects. 
%AMUSE is ideally suited to model these problems. The framework forces 
%the implementation handling of the physics for the coupling into the 
%controlling layer. The researcher can treat a code as a \emph{black 
%box} and focus on the physics of coupling. In multi-scale and 
%multi-domain problems, several codes will have to be coupled and by 
%being able to change and control each of these, we can ensure the 
%physics is implemented correctly. Also, like codes, we can experiment 
%with different coupling methods to see how these influence the evolving 
%system.

\section*{Acknowledgment}

This work was initiated during the 2013 "Multiscale Modelling and
Computing" workshop at the Lorentz Center, Leiden, The Netherlands,
and is supported by the Netherlands Research Council NWO (grants
\#643.200.503, \#639.073.803 and \#614.061.608) and by the Netherlands
Research School for Astronomy (NOVA).  This research has partially
been funded by the Interuniversity Attraction Poles Programme
initiated by the Belgian Science Policy Office (IAP P7/08 CHARM).

\bibliographystyle{plain}
\bibliography{./mdms}

\end{document}